\documentclass[%
 aip,
 jmp,%
 amsmath,amssymb,
preprint,%
]{revtex4-1}
\usepackage[caption=false]{subfig}%
\usepackage[utf8]{inputenc} %
\usepackage{graphicx}
\usepackage{dcolumn}
\usepackage{bm}
\usepackage{epstopdf}
\usepackage{amssymb}
\usepackage{color}
\usepackage[colorlinks=true, letterpaper=true, pdfstartview=FitV, linkcolor=blue, citecolor=blue, urlcolor=blue]{hyperref}
\usepackage{lineno}

\begin{document}

\title{First-principles calculations on the mechanical, electronic, magnetic and optical properties of two-dimensional Janus  Cr$_{2}$TeX (X= P, As, Sb) monolayers}
\author{Qiuyue Ma}
\author{Wenhui Wan}
\author{Yanfeng Ge}
\author{Yingmei Li}
\author{Yong Liu}\email{yongliu@ysu.edu.cn}
\affiliation{State Key Laboratory of Metastable Materials Science and Technology \& Key Laboratory for Microstructural Material Physics of Hebei Province, School of Science, Yanshan University, Qinhuangdao 066004, China }

\begin{abstract}
Janus materials possess extraordinary physical, chemical, and  mechanical properties caused by symmetry breaking. Here, the mechanic properties, electronic structure, magnetic properties, and optical properties of Janus Cr$_2$TeX (X= P, As, Sb) monolayers are systematically investigated by the density functional theory. Janus Cr$_2$TeP, Cr$_2$TeAs, and Cr$_2$TeSb are intrinsic ferromagnetic (FM) half-metals with wide spin and half-metallic gaps. The Curie temperature ( \emph{T}$_c$ ) of these monolayers are about 583, 608, and 597 K, respectively by Monte Carlo simulations based on the Heisenberg model. Additionally, it has been found that Cr$_2$TeX (X= P, As, Sb) monolayers still exhibit FM half-metallic properties under biaxial strain ranging from -6\% to 6\%. At last, the Cr$_2$TeP monolayer has a higher absorption coefficient than the Cr$_2$TeAs and Cr$_2$TeSb monolayers in the visible region. The results predict that Janus Cr$_2$TeX (X= P, As, Sb) monolayers with novel properties have good potential for applications in future nanodevices.
\end{abstract}

\maketitle



The discovery of two-dimensional (2D) honeycomb structures of graphene is considered one of the most important discoveries in the field of physics and materials science~\cite{1K. S.-Science-2004}. The unique outstanding physical and chemical properties of graphene have led to an upsurge in the study of 2D materials. Then, a large number of 2D layered materials have been reported, including hexagonal boron nitride~\cite{2K. Novoselov-Proc.-2004,3C. Jin-Phys. Rev. Lett.-2009}, transition metal dichalcogenides (TMDs)~\cite{4M. Chhowalla-Nat. Chem.-2013,5K. F.-Phys. Rev. Lett.-2010,6A. Splendiani-Nano Lett.-2010}, silicene~\cite{9B. Lalmi-Appl. Phys. Lett.-2010,10S.-C-Phys. Rev. Lett.-2014,11M. Yang-Front. Phys.-2015} have shown great potential for various applications. 2D Janus monolayers show novel physical and chemical properties, such as piezoelectric polarization~\cite{12C. Zhang-Nano Lett.-2019}, Rashba effect~\cite{9Q.-F.-Phys. Rev. B-2017,10T. Hu-Phys. Rev. B-2018}, and catalytic performance~\cite{13D. Er-Nano Lett.-2018,14X. Ma-J. Mater. Chem. A-2018}, which open up new possibilities in the field of 2D materials. The Janus In$_2$SeTe is synthesized by substituting one layer of Se atoms with Te atoms to break the inversion symmetry of InSe, which possesses transport characteristics that are superior to InSe monolayer~\cite{22W. Wan-J. Phys-2019}. Similarly, the Janus In$_2$SSe monolayer has an indirect-direct bandgap transition due to broken vertical symmetry~\cite{27A. Kandemir-Phys. Rev. B-2018}. 2D Janus PdXO (X= S, Se, Te) monolayers with high electron mobility have good potential for applications in future nanodevices~\cite{23T. V.Vu-RSC Adv.-2022}. For Janus TMDs materials, the MXY (M = Mo, W; X, Y = S, Se, Te; X $\neq$ Y) monolayers show excellent intrinsic dipole and piezoelectric effects~\cite{12L. Dong-ACS Nano-2017}.

Recently, many 2D magnetic materials have been predicted, such as  CrI$_3$ ~\cite{41B. Huang-Nature-2017}, Cr$_2$Ge$_2$Te$_6$ ~\cite{42C. Gong-Nature-2017}, V$_3$X$_8$~\cite{H. Xiao-p-2019}, Cr$_2$NX$_2$ (X = O, F, OH)~\cite{Q-a-2021}, and M$_2$SeTe (X= Ga, In)~\cite{Y. Guo-a-2017}. Meanwhile, recent works also confirm that many Janus magnetic materials have excellent properties, such as Cr$_2$I$_3$X$_3$ (X = F, Cl, Br)~\cite{23W-m-2022}, FeXY (X, Y = Cl, Br, and I, X $\neq$ Y)~\cite{28R. Li-Nature Nanotechnology-2017}, XGaInY (X, Y;= S, Se and Te)~\cite{I. AhMaD-RSC Adv.-2021}. 2D Janus monolayers possess extraordinary physical and chemical characteristics, which have potential applications in 2D nanoscale spintronic devices. Half-metals with one spin channel conducting and the other semiconducting filter the current into a single spin channel for realizing pure spin transport, generation, and injection. The FeCl$_2$ is experimentally known to exist in a monolayer form and is a classical material with half-metallic properties ~\cite{E. Torun-Appl. Phys. Lett.-2015}, but its Curie temperature (\emph{T}$_c$) is only 17 K, which greatly affects its application in spintronics due to its relatively low \emph{T}$_c$. In recent years, it has been recognized that the half-metallic materials have exposed high \emph{T}$_c$ as a result of the strong exchange interaction driven by charge carriers~\cite{C. Zener-Phys. Rev.-1951}. Theoretical studies have predicted many 2D half-metals with high \emph{T}$_c$, Fe$_2$Si for 780 K~\cite{S-N-2017}, MnAsS$_4$ for 740 K~\cite{T. Hu-J. Phys-2020}, MnP and MnAs for 495 K and 711 K~\cite{B. Wang-Nanoscale-2019}, and Mn$_2$AsP monolayer for 557 K~\cite{H. Zeng-J. Mater. Sci.-2021}. 2D half-metals, which have abundant charge carriers, are expected to be much higher \emph{T}$_c$ values than 2D magnetic semiconductors. The outstanding attributes of half-metals are potentially promising for the fabrication of future nanodevices.

In this work, the structural characteristics, mechanical, electronic properties, magnetic properties, and optical properties of Janus Cr$_2$TeX (X= P, As, Sb) monolayers have been studied based on the first-principles calculations. The results show that the Janus Cr$_2$TeX (X= P, As, Sb) monolayers are intrinsic ferromagnetic half-metals with wide half-metallic gaps and spin gaps. The predicted \emph{T}$_c$ of Janus Cr$_2$TeX (X= P, As, Sb) monolayers reached up to 583, 608, and 597 K. Janus Cr$_2$TeX (X= P, As, Sb) monolayers still retain their half-metallic properties and the FM natures are robust against biaxial strain in the range of -6\% to 6\%. At last, we displayed their optical properties, the Cr$_2$TeSb has a higher absorption coefficient and lower energy-loss coefficient in the ultraviolet region. Our calculations indicate that Janus Cr$_2$TeX (X= P, As, Sb) monolayers are potentially promising for spintronic devices.

The present calculations were performed by adopting the Vienna ab initio simulation package (VASP) based on the density functional theory (DFT)~\cite{27P.-Phys. Rev. B-1994,28G. Kresse-American Physical Society-1996,29G. Kresse-Computational Materials Science-1996}. The generalized gradient approximation (GGA) functional of Perdew, Burke, and Ernzerhof (PBE) was used to investigate the exchange-correlation
function~\cite{30J. P. Perdew-Phys. Rev. Lett.-1996}. We used the spin-dependent GGA plus Hubbard U to deal with the strongly correlated interactions of the transition metal Cr element, the Hubbard U term of 3 eV was used for Cr~\cite{32A. I. Liechtenstein-American Physical Society-1995}. The plane-wave cutoff energy was chosen to be 500 eV. Monkhorst-Pack special k-point mesh is $9\times9\times1$ for the Brillouin zone integration~\cite{31H. J. Monkhorst-Phys. Rev. B-1976}. The convergence criteria for energy and force during the relaxation of the structures were set to 10$^{-6}$ eV and 0.01 eV/{\AA}. The vertical vacuum spacing of 20 {\AA} was used to eliminate interactions between images. Phonon dispersions of the studied materials were obtained using the phonopy code based on the density functional perturbation theory (DFPT)~\cite{33S. Baroni-Rev. Mod. Phys.-2001}.

\begin{figure}[t!hp]
\centerline{\includegraphics[width=0.95\textwidth]{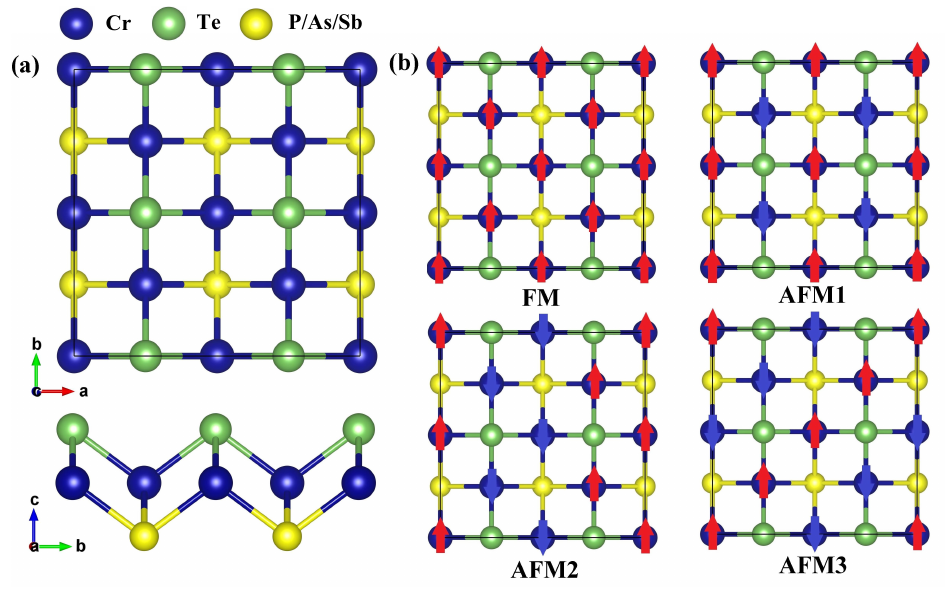}}
\caption{(a)Top and side views of Janus Cr$_2$TeX (X= P, As, Sb) monolayers. (b) One FM and three AFM configurations of Janus Cr$_2$TeX (X= P, As, Sb) monolayers.
\label{fig:1}}
\end{figure}


The crystal structure of Janus Cr$_2$TeX (X= P, As, Sb) monolayers is shown in Fig.1(a). One Cr layer is sandwiched between the X layer and Te layer. The lattice constants of Janus Cr$_2$TeX monolayers increase from 4.50 of Cr$_2$TeP to 4.65 {\AA} of Cr$_2$TeSb due to the increase in the size of the X atom. The calculated structure parameters of Janus Cr$_2$TeX (X= P, As, Sb) monolayers are summarized in Table~\ref{table:elastic}. As shown in Fig.1(b), four typical magnetically coupled configurations in the 2 $\times$ 1 $\times$ 1 supercell are considered to study the ground state of the Cr$_2$TeX ( X = P, As, Sb) monolayers. Four spin configurations confirm that the FM coupling is more energetically stable than antiferromagnetic(AFM) coupling. Additionally, the non-magnetic state can be neglected due to the energy difference between the non-magnetic state and the magnetic state is too huge. In general, the magnetic state of the crystal structure is determined by the competition between two mechanisms. One is the direct AFM interaction between the nearest neighboring Cr atoms, the other is  superexchange interaction among Cr-Te-Cr or Cr-X-Cr. In monolayers Cr$_2$TeX (X= P, As, Sb), the Cr-Te-Cr or Cr-X-Cr bond angles are all closer to 90$^\circ$, which usually favor FM ordering according to the Goodenough-Kanamori-Anderson (GKA) rules ~\cite{36J. B-Phys. Rev.-1955,37J. Kanamori-Journal of Applied Physics-1960}. The superexchange FM dominates the total exchange interaction in Janus Cr$_2$TeX (X= P, As, Sb) monolayers due to the large Cr-Cr distance, which weakens the AFM direct exchange interaction. As a result, Janus Cr$_2$TeX (X= P, As, Sb) monolayers  all adopt the FM ground state.

\begin{table*}[!t]\small
\centering \caption{Lattice constants: a ({\AA}), bond length: d$_{Cr-Cr}$, d$_{Cr-X}$, and d$_{Cr-Te}$ ({\AA}), the angle of Cr-Te/X-Cr : $\theta${$_1$}/$\theta${$_2$} ($^\circ$), spin gap (eV), half-metallic gap (eV), exchange constants: J{$_1$} and J{$_2$} (meV per Cr), magnetic anisotropy energy: MAE ($\mu$eV per Cr), and Curie temperature: \emph{T}$_c$ (K) of Janus Cr$_2$TeX (X= P, As, Sb) monolayers.}
\renewcommand\arraystretch{1.6}  
\begin{tabular}{cccccccccccccccccc}
  \hline
  System  \quad &  a  \qquad &  d$_{Cr-Cr}$ \qquad &  d$_{Cr-X}$  \qquad &  d$_{Cr-Te}$\qquad &  $\theta${$_1$}  \quad &  $\theta${$_2$}   \quad &   spin gap  \quad &  half-metallic gap  \quad & J{$_1$} \quad & J{$_2$}\quad & MAE \quad & \emph{T}$_c$  \quad   \\
  \hline
  Cr$_2$TeP  & 4.50 &3.18 &2.43 &2.75& 70.63 & 82.00 & 2.84 & 1.25 & 22.83 & 14.31  &185.02& 583    \\
  Cr$_2$TeAs  & 4.54 & 3.21 & 2.53 & 2.75& 71.31& 78.64& 2.96 &1.01 & 25.10 &16.90 &672.33 &608 &  \\
  Cr$_2$TeSb  & 4.65 & 3.29 & 2.75 & 2.76 & 73.19& 73.36& 2.64 &0.56 &24.42 & 9.28 &371.43&597 &  \\
  \hline
  \label{table:elastic}
\end{tabular}
\end{table*}

We then evaluated the mechanical stability of Cr$_2$TeX (X= P, As, Sb) by calculating their elastic coefficients C$_{ij}$. As shown in Table~\ref{tablep}, C$_{ij}$ of Cr$_2$TeX decrease with increasing atomic size of X element. The elastic constants of Cr$_2$TeX (X= P, As, Sb) meet the Born criterion of stability (C$_{11}{>}$0, C$_{22}{>}$0 and C$_{11}$-C$_{12}{>}$0)~\cite{34Z.-Phys. Rev. B-2007}, indicating that the Cr$_2$TeP, Cr$_2$TeAs, and Cr$_2$TeSb are mechanically stable. In addition, the \emph{Y}$_{2D}$ of Janus Cr$_2$TeP, Cr$_2$TeAs, and Cr$_2$TeSb monolayers were calculated to be 22.6 N/m, 16.67 N/m, and 12.88 N/m as shown in Table~\ref{tablep}. The \emph{Y}$_{2D}$ of Cr$_2$TeX is smaller than that of other 2D layered nanostructures, such as graphene (340 N/m)~\cite{D. G. Papageorgiou-Prog. Mater. Sci.-2017} and MoS$_2$ (130 N/m)~\cite{R. C-Phys. Rev. B-2013}. The small \emph{Y}$_{2D}$ proves they possess high flexibility and can sustain large strain. Furthermore, the dynamic stability was predicted by phonon dispersion calculations. The phonon spectrum of Cr$_2$TeX (X= P, As, Sb) without negative frequencies proves that Cr$_2$TeP, Cr$_2$TeAs and Cr$_2$TeSb are dynamically stable [Fig.S1].

\begin{table}[h]
\begin{ruledtabular}
\caption{Elastic constants C$_{11}$, C$_{12}$ , and C$_{66}$ (N/m), Young's modulus \emph{Y}$_{2D}$ (N/m), and Poisson's ratio $\nu$ of Janus Cr$_2$TeX (X= P, As, Sb) monolayers in FM state.}\label{tablep}
\begin{tabular}{lccccc}
     & C$_{11}$ & C$_{12}$ & C$_{66}$ & \emph{Y}$_{2D}$  & $\nu$ \\
   \hline

Cr$_2$TeP & 24.47 & 7.69 & 8.39 & 22.06 & 0.31 \\

Cr$_2$TeAs & 17.20 & 3.01 & 7.10 & 16.67 & 0.17 \\

Cr$_2$TeSb & 13.25 & 2.21 & 5.52 & 12.88 & 0.16 \\
\end{tabular}
\end{ruledtabular}
\end{table}

\begin{figure}[htp]
\centerline{\includegraphics[width=0.95\textwidth]{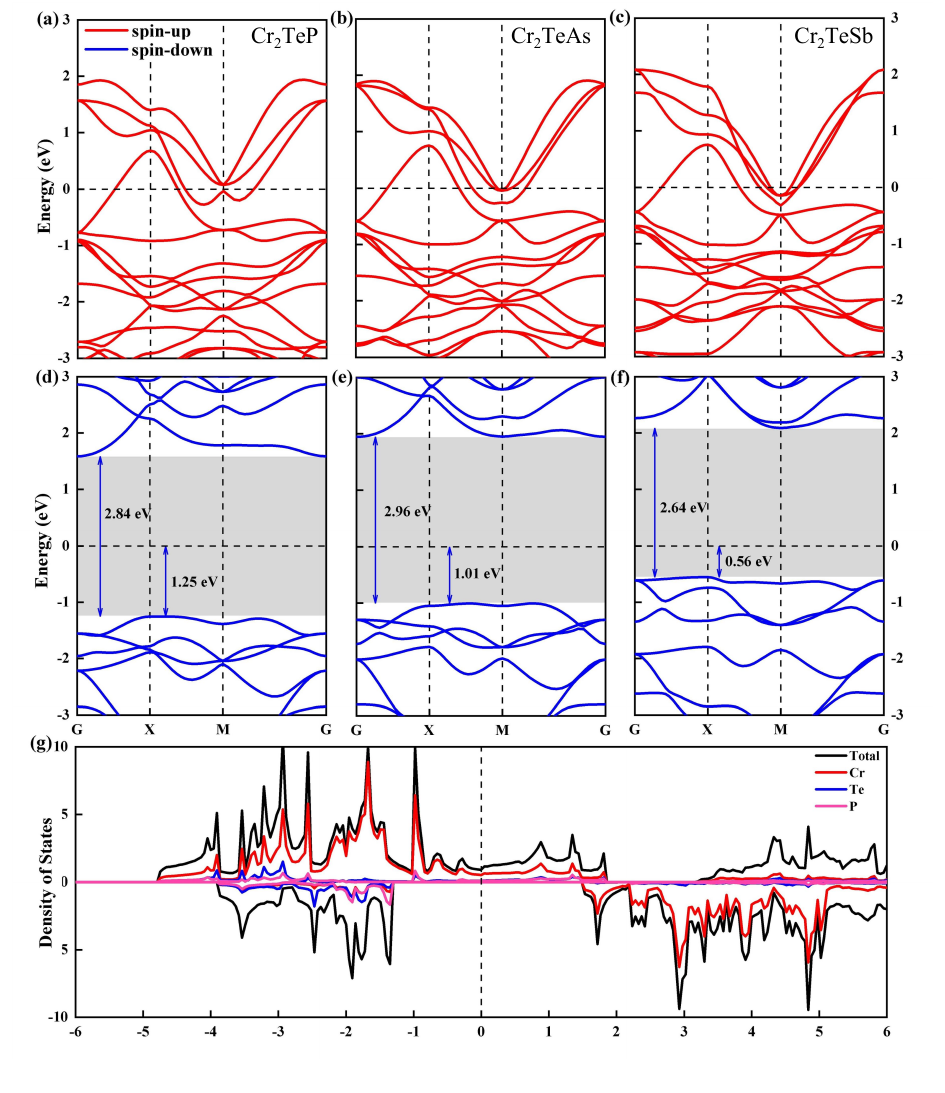}}
\caption{The electronic band structures of Cr$_2$TeP (a,d), Cr$_2$TeAs (b,e), and Cr$_2$TeSb (c,f). The total density of states (TDOS) and projected density of states (PDOS) of Cr$_2$TeP monolayer (g).
\label{fig:2}}
\end{figure}

Fig.2(a)-(f) shows the band structures of Cr$_2$TeX at the PBE level. Notably, the spin-up bands cross the Fermi level, while the spin-down channel acts as a semiconductor, indicating that they are intrinsic half-metals. The calculated electronic band structures of the Cr$_2$TeX exhibit half-metallic gaps ranging from 1.25 eV of Cr$_2$TeP to 0.56 eV of Cr$_2$TeSb, which can effectively prevent thermally excited spin-flip transitions. The band gaps on the spin-down channel were 2.84, 2.96, and 2.64 eV for Cr$_2$TeP, Cr$_2$TeAs, and Cr$_2$TeSb at the PBE levels, respectively, which is beneficial for the spin-polarized carrier injection and detection. Because the PBE functional tends to underestimate the band gap, the HSE06 functional was used to correct the band structures, which showed the band gaps in the spin-down channel increased to 3.70, 3.84, and 3.53 eV for the Cr$_2$TeP, Cr$_2$TeAs, and Cr$_2$TeSb monolayers, respectively. In the partial density of states (PDOS) of Cr$_2$TeX [Fig.2(g) and Fig.S2], the spin-up electronic states near the Fermi level are mainly contributed by Cr atoms. For the spin-down electronic states, the conduction band minimum is mainly contributed by Cr atoms, while the Cr, X, and Te atoms all contribute to the valence band maximum.

To estimate the \emph{T}$_c$ of Janus Cr$_2$TeX (X= P, As, Sb) monolayers, we performed Monte carlo simulations based on the Heisenberg model with the Hamiltonian of:

\begin{equation}
H =  - \sum\limits_{ < ij > } {{J_{1}}{S_i}{S_j}}- \sum\limits_{ < ik > } {{J_{2}}{S_i}{S_k}}- A{{S_i^Z}{S_i^Z}}
\end{equation}

Where J$_1$ and J$_2$ are the exchange constants, S$_i$ and S$_j$ are the spin operators on sites i and j, respectively. $S_i^Z$ is the spin parallel to the Z direction. And A is the magnetic anisotropy energy parameter (MAE). J$_1$ and J$_2$ can be obtained by equations:

\begin{equation}
E_{FM} = E_{0} - 2J_{1}{S^2}- 2{J_2}{S^2}- A{S^2}   \\
\end{equation}

\begin{equation}
E_{AFM1} = E_{0} + 2J_{1}{S^2}- 2J_{2}{S^2}- A{S^2} \\
\end{equation}

\begin{equation}
E_{AFM3} = E_{0} + 2J_{2}{S^2}- A{S^2}             \\
\end{equation}

The calculated J$_1$, J$_2$, and MAE are listed in Table~\ref{table:elastic}. As shown in Fig.3, \emph{T}$_c$  are estimated to be 583, 608, and 597 K for the Cr$_2$TeP, Cr$_2$TeAs, and Cr$_2$TeSb, respectively, much higher than those in the 2D FM half-metals reported early, e.g., FeClBr (28 K), FeClI (21 K), FeBrI (29 K)~\cite{28R. Li-Nature Nanotechnology-2017}, ScCl (185 K)~\cite{41B. Wang-Nanoscale Horiz.-2018}, and VSeTe (350 K)~\cite{z-Nanoscale-2020}. The Janus Cr$_2$TeP, Cr$_2$TeAs, and Cr$_2$TeSb with high \emph{T}$_c$ are benefit for possible applications in spintronic devices.

\begin{figure}[htp]
\centerline{\includegraphics[width=0.95\textwidth]{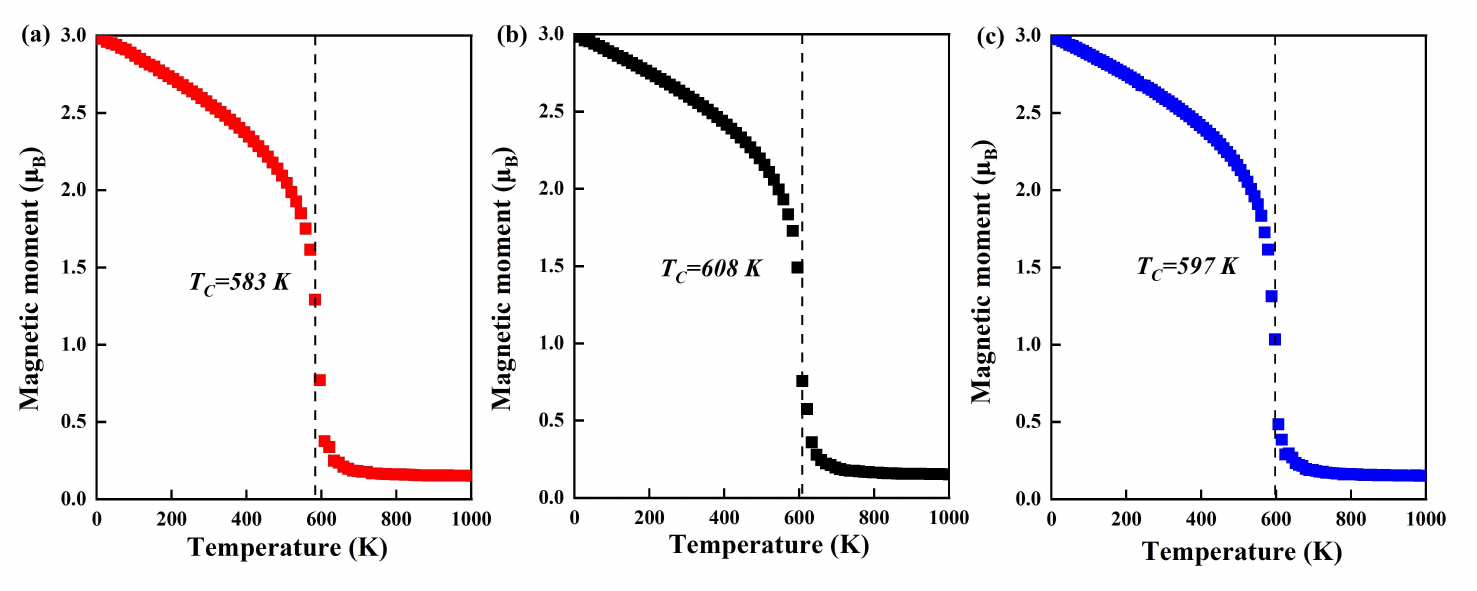}}
\caption{On-site magnetic moment of Cr atoms versus temperature in Cr$_2$TeP (a), Cr$_2$TeAs (b), and Cr$_2$TeSb (c).
\label{fig:3}}
\end{figure}

Then, we study the effects of biaxial strain on the electronic and magnetic properties of Cr$_2$TeX (X= P, As, Sb) monolayers. As shown in Fig.4(a), the energy difference between the FM and AFM configurations under the biaxial strain indicates that no phase transition occurs during the process of biaxial strain and FM is always the ground state. And the energy difference of Cr$_2$TeAs and Cr$_2$TeSb increases with the increase in tensile strain and decreases with the compressive strain. The half-metallic gaps in the Janus of Cr$_2$TeP and Cr$_2$TeAs monolayers monotonously decrease as biaxial strain increases from -6\% to 6\%[Fig.4(b)]. Meanwhile, the half-metallic gap of Cr$_2$TeP reaches a maximum value under 2\% strain. When biaxial strain of -2\%, 0\%, and 2\% are applied for Cr$_2$TeP, Cr$_2$TeAs, and Cr$_2$TeSb, respectively, the spin gaps reach the maximum value of 2.98, 2.96, and 2.83 eV [Fig.4(c)].

\begin{figure}[t!hp]
\centerline{\includegraphics[width=0.95\textwidth]{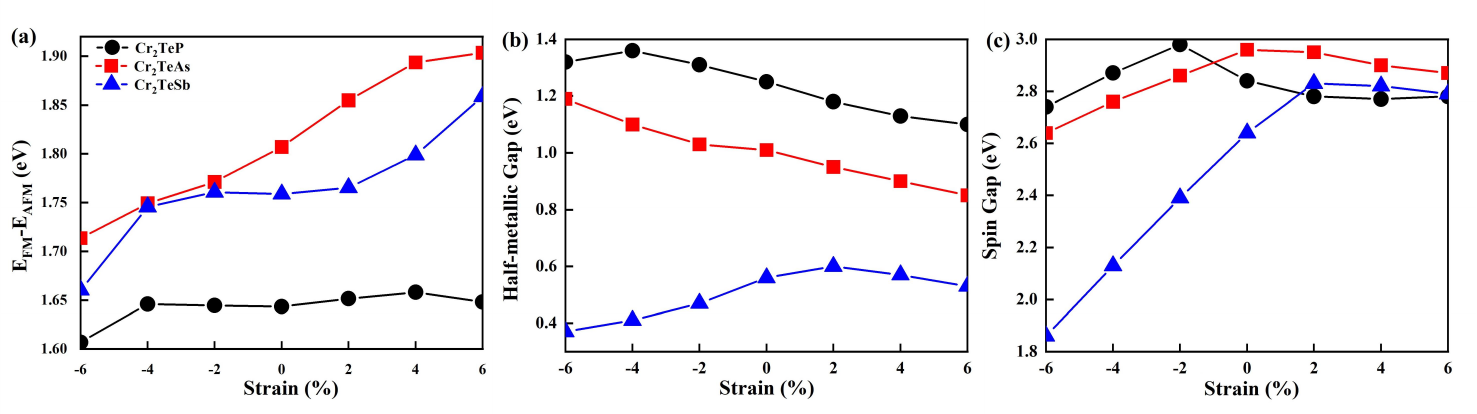}}
\caption{Energy difference (a) between the FM and AFM phases, variation of the half-metallic gap (b) and spin gap (c) under biaxial strain for Janus Cr$_2$TeP, Cr$_2$TeAs, and Cr$_2$TeSb.
\label{fig:4}}
\end{figure}

At last, we investigated the optical properties of Cr$_2$TeX (X= P, As, Sb) monolayers by including the photon absorption coefficient $\alpha$($\omega$), energy-loss coefficient L($\omega$), reflectivity R($\omega$), and refraction coefficient n($\omega$). As shown in Fig.5(a), in the energy range of 0-5 eV, the overall absorption coefficient of Cr$_2$TeX monolayers show a gradually increasing trend. The absorption coefficient in the visible range (1.3-3.5 eV) is higher in Cr$_2$TeP monolayer than in the Cr$_2$TeAs and Cr$_2$TeSb monolayers as a result of the higher $\alpha$($\omega$) of the Cr$_2$TeP monolayer in this range. In the ultraviolet region (3.5-5 eV), Cr$_2$TeSb shows a higher $\alpha$($\omega$). Fig.5(b) shows that L($\omega$) of the Cr$_2$TeSb is fairly low in the visible range (1.3-3.5 eV), and its L($\omega$) is also significantly lower than that of the Cr$_2$TeAs and Cr$_2$TeSb monolayers in the ultraviolet region (3.5-5 eV). The lower L($\omega$) indicates that the Cr$_2$TeSb monolayer has favorable photon transmittance. The refraction coefficient can be used to characterize the attenuation of light energy. We can see that Cr$_2$TeSb monolayer has a higher reflectivity R($\omega$) in the infrared region (0.6-1.3 eV)[Fig.~5(c)(d)]. The higher the refraction coefficient, the more the incident, and the higher the ability of light to be refracted. The overall trend of the refraction coefficient of Cr$_2$TeX monolayers is gradually decreasing. In the infrared region of photon energy 0-1.3 eV, the refraction coefficient decays rapidly, and the Cr$_2$TeP has has a higher n($\omega$) than the other two monolayers. In the visible range of 1.3-3.5 eV, the refraction coefficient tends to weaken with the decay of the incident light energy, and the decay tends to level off in the ultraviolet region at 3.5-5.0 eV, and the difference in refractive power with Cr$_2$TeX in the ultraviolet region is not significant.

\begin{figure}[htp]
\centerline{\includegraphics[width=0.95\textwidth]{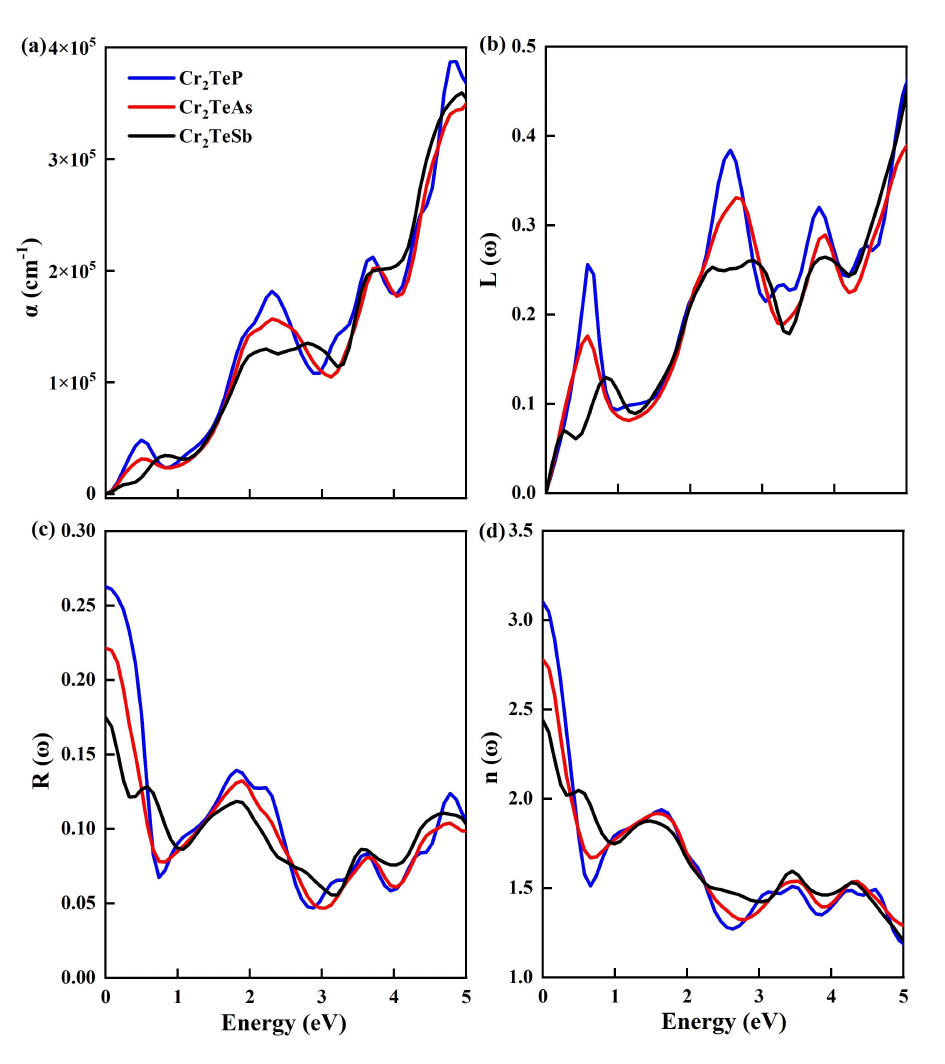}}
\caption{Optical property diagram of Cr$_2$TeX (X= P, As, Sb) monolayers: photon absorption coefficient (a), energy-loss coefficient (b), reflectivity (c), and refraction coefficient (d).
\label{fig:5}}
\end{figure}

In conclusion, the structural, mechanical, electronic, magnetic, and optical properties of Janus Cr$_2$TeX (X= P, As, Sb) monolayers have been studied by means of DFT calculations. The calculated results demonstrated that all three structures of Cr$_2$TeX were found to be mechanically and dynamically stable. We have predicted that the Janus Cr$_2$TeX monolayers are large band gap FM half metals. Meanwhile, we further reveal that Cr$_2$TeX (X= P, As, Sb) show a high \emph{T}$_c$ about 583, 603, and 597 K, respectly. The FM half-metallic characteristics can exist stably in a large strain range from -6\% to 6\%. The optical properties show that Cr$_2$TeSb has higher absorption coefficient and lower energy-loss coefficient in the ultraviolet region. The present results indicate that these Janus Cr$_2$TeX (X= P, As, Sb) monolayers have potential functional materials for spintronic applications and enrich the 2D Janus half-metallic material library.

Supplementary Material includes 2 figures.

This work was supported by the Innovation Capability Improvement Project of Hebei province (No. 22567605H), the Natural Science Foundation of Hebei Province of China (No. A2022203006), the Science and Technology Project of Hebei Education Department (No. BJK2022002 and QN2023177). The numerical calculations in
this paper have been done on the supercomputing system in the High Performance Computing Center of Yanshan University.

\section{DATA AVAILABILITY}
The data that support the findings of this study are available from the corresponding author upon reasonable request.


\bibliography{apssamp}

\end{document}